\documentstyle[aps, manuscript, epsf]{revtex}

\begin{document}

\title{
Thermo-refractive noise in gravitational wave antennae}

\author{V. B.\ Braginsky,\ M. L. Gorodetsky and S. P.\ Vyatchanin, \\[3mm]
}
\address{
Faculty of Physics, Moscow State University, Moscow 119899, Russia \\
	e-mail: vyat@hbar.phys.msu.su
}
\date{\today}

\maketitle

\begin{abstract}
Thermodynamical fluctuations of temperature in mirrors of gravitational wave
antennae may be transformed into additional noise not only through thermal expansion
coefficient \cite{bgv} but also through temperature dependence of refraction
index. The intensity of this noise is comparable with other known noises 
and must be taken into account in future steps of the antennas. 
\end{abstract}

\section{Introduction}

We have shown in our previous article \cite{bgv} that thermodynamical
fluctuations of temperature in mirrors (test masses) of LIGO-type gravitational wave
antenna \cite{abr1,abr2} are transformed due to 
the thermal expansion coefficient $\alpha=(1/l)(dl/dT)$ into additional 
(thermoelastic) noise which may be a serious
"barrier" limiting sensitivity. This noise is caused in fact
by random fluctuations of the coordinate averaged over the mirror's
surface. The spectral density of this random coordinate displacement may 
be presented for infinite test mass in the following form
\footnote{
This result was refined for the case of finite test masses by
Yu. T. Liu and K. S. Thorne \cite {kip}. However the difference from
our calculation is only several percents for the planned sizes of test
masses, and hence we use here much more compact expression (\ref{tespot}).
}:
\begin{eqnarray}
\label{tespot}
S^{\rm TE}_{x,\,\alpha}(\omega)& = &\frac{8}{\sqrt{2\pi}}\,
	\frac{ \kappa T^2 \alpha^2(1+\sigma)^2\,\lambda^* }
		{(\rho C)^2 \,r_0^3\, \omega^2}.
\end{eqnarray}
Here $\kappa$ is the Boltzmann constant, $T$ is temperature, $\sigma$ is Poison 
ratio, $\lambda^*$ is thermal conductivity, $\rho$ is density 
and $C$ is specific heat capacity, $r_0$ is the radius of the spot of laser
beam over which the averaging of the fluctuations is performed. This noise
is of nonlinear origin as the nonzero value of $\alpha$ is due to the anharmonisity
of the lattice.

The goal of this article is to present the results of the analysis of another
additional (and also of nonlinear origin) effect which may be comparable with
other known noise mechanisms limiting the sensitivity. Qualitatively this
effect is may be understood in the following way. The laser beam ``extracts''
the information not only about the change of the length of the antenna produced
by gravitational wave but also about the fluctuations of position of mirrors'
surfaces averaged over the beam spot. These fluctuations lead to phase noise
in the reflected optical field. However the phase noise may be produced by
another effect. High reflectivity of the mirrors is provided by multilayer
coatings which consist of alternating sequences of quarter-wavelength dielectric layers
having refraction indices $n_1$ and $n_2$. The most frequently used pairs of
layers are $TiO_2-SiO_2$ and  $Ta_2O_5-SiO_2$. \marginpar{*} 
While reflecting the optical wave "penetrates"
in the coating on certain depth. This depth is of the order of the optical
thickness of one pair of layers $l<1\mu $. If the values of $n_1$ and $n_2$
depend on temperature $T$ (thermo-refractive factor $\beta=dn/dT$ is nonzero)
then thermodynamical fluctuations of temperature lead to fluctuations of
optical thickness of these layers and hence to the phase noise in the
reflected wave. Though the thickness $l$ of the working layer is small,
the coefficient $\beta$ is usually significantly larger than $\alpha$
(both have the same dimensions). For fused silica ($SiO_2$)
$\alpha=5\times 10^{-7}\ K^{-1}$
and $\beta=1.45\times 10^{-5}\ K^{-1}$ (i.e. 30 times larger than $\alpha$).
This phase noise may be evidently easily recalculated in terms of equivalent
fluctuations of the surface and consequently compared with the spectral
sensitivity of the antenna.

We have analyzed also the photo-thermal refractive shot noise: due to random
absorption of optical photons, the random fluctuations of temperature in the
surface layer of the mirror take place, producing fluctuations of refractive
indices of the coating and therefore phase fluctuations of reflected light
wave (this effect is similar to photo-thermal shot noise, analyzed in
\cite{bgv}). However, this effect is  numerically much smaller than
thermo-refractive noise --- that is why we do not present here the detailed
analysis of it.

\section{Thermo-refractive noise}\label{sec:2}

The theory of reflection of light from multilayer dielectrical coating is
well known (see for example \cite{solimeno}). Using traditional approach we may
recalculate the phase shift $\delta \phi$ into equivalent displacement $\delta x$ 
of mirror (see Appendix \ref{A}):
\begin{eqnarray}
\label{deltax}
\delta x = \frac{\lambda}{4\pi}\delta\phi=- \bar u\lambda\beta_{\rm eff},\\
\label{betaeff}
\beta_{\rm eff}=\frac{n^2_2\beta_1+n^2_1\beta_2}{4(n_1^2-n^2_2)}.
\end{eqnarray}
Here $\bar u$ is the fluctuation of averaged temperature, $\beta_1=dn_1/dT$, $\beta_2=dn_2/dT$ . 
It is important to note, that  effective coating thickness is much smaller
than the characteristic length of diffusive heat transfer:
$l\ll a / \sqrt{\omega} $ ($a$ is temperature conductivity, $\omega$ is the
frequency of observation which is of order $\sim 100Hz$ for laser
gravitational wave antenna). Therefore we may consider in our calculations
that fluctuations of temperature  are correlated in the layers.

To calculate thermodynamical fluctuations of temperature $u(\vec r,t)$ 
in the surface layers
we use Langevin approach and introduce fluctuational thermal sources
$F(\vec r,t)$ added to the right part of the equation of
thermal conductivity:
\begin{eqnarray}
\label{TDFT}
\frac{\partial u}{\partial t} - a^2\Delta u = F(\vec r, t),\qquad
a^2=\frac{\lambda^*}{\rho C}.
\end{eqnarray}

This approach was described and verified in \cite{bgv}
(see all the details over there). 
As in \cite{bgv} we replace  the
mirror by half-space: $-\infty<x<\infty,\ -\infty<y<\infty,\ 
0\le z<\infty$ with the boundary condition of thermo-isolation on
surface $z=0$. We may now calculate the spectrum of temperature fluctuations:
\begin{eqnarray} \label{u}
u(\vec r, t)&=&\int^{\infty}_{-\infty}\frac{d\vec k\, d\omega}{(2\pi)^4}\ u(\omega, \vec k)
	e^{i\omega t +i\vec k\vec r},\label{ur}\\
u(\omega, \vec k)&=&\frac{F(\vec k,\omega)}{a^2(\vec k )^2+ i\omega}
	\label{uk},\\
\label{corrFk}
\langle F(\vec k,\omega)F^*(\vec k_1,\omega_1)\rangle&=&
	 \frac{2\kappa T^2 \lambda^*}{ (\rho C)^2 }\,
	(2\pi)^4 |\vec k|^2\,
	\delta(\omega-\omega_1)\times\\
&\times&
	\delta (k_x-k_{x1})\,\delta (k_y-k_{y1})\times\nonumber\\
&\times&
	\big[\,\delta (k_z-k_{z1})+\delta (k_z+k_{z1})\,\big].\nonumber 
\end{eqnarray}

The thermodynamical fluctuations of temperature $\bar u$ averaged over the 
volume $V=\pi r_0^2l$ may be presented in the following form:
\begin{eqnarray}
\bar u&=&\frac{1}{\pi r_0^2 l}
	\int^{\infty}_{-\infty}dx dy \int^{\infty}_{0}dz \ u(\vec r, t)\
	e^{-(x^2+y^2)/r_0^2}\,e^{-z/l}= \nonumber \\
\label{baru1}
&=&\int^{\infty}_{-\infty}\frac{d\vec k\, d\omega}{(2\pi)^4}\
	\frac{F(\vec k,\omega)e^{i\omega t} }{a^2|\vec k |^2+ i\omega}\
	e^{-(k_x^2+k_y^2)r_0^2/4}\frac{1}{1-ik_zl},
\end{eqnarray}

From this expression and from (\ref{corrFk}) we find immediately the 
spectral density
$S_u(\omega)$ of fluctuations of the averaged temperature: 
\begin{eqnarray}
S_u(\omega)    &=&2\times \frac {2\kappa T^2\,\lambda^*}{(\rho C)^2}\, 
	 \int_0^\infty \frac{2\pi\,k_\bot dk_\bot }{(2\pi)^2} 
	\int^{\infty}_{-\infty}\frac{dk_z}{2\pi}\times \nonumber \\
&\times& \frac{k_z^2+k_\bot^2}{a^4(k_z^2+k_\bot^2 )^2+ \omega^2}\
	e^{-k_\bot^2r_0^2/2 }\frac{(1+1)}{1+k^2_z l^2}=\nonumber\\
&\simeq &
	\frac {\sqrt{2}\,\kappa T^2}{\pi\,r_0^2 \sqrt{\omega\lambda^*\rho C}}
\end{eqnarray}
Here $k_\bot^2=k_x^2+k_y^2$.
The first term $2$  appears because as in \cite{bgv} we use 
``one-sided'' spectral density, defined only for positive
frequencies, which is connected with the correlation function
$\langle  u(t)u(t+\tau)\rangle$ by the formula
$
S_u(\omega)=2\int_{-\infty}^{\infty}d\tau\,
        \langle u(t)u(t+\tau)\rangle\,\cos (\omega\tau).
$
The term $(1+1)$ appears due to two $\delta$-functions in square brackets
in  (\ref{corrFk}). For the frequency of observation $\omega \simeq 2\pi\times 100\ \mbox{s}^{-1}$
characteristic length $a/\sqrt{\omega}\simeq 50\ \mu$ 
(we used for the estimates constants for fused silica), so that $l\ll a/\sqrt{\omega}\ll r_0$.
Taking into account that $k_\bot \simeq 1/r_0\ll \sqrt{\omega}/a$
we may consider the first denominator as constant while integrating 
over $k_\bot$. In the same way $k_z \simeq 1/l \gg \sqrt{\omega}/a$
and while integrating over $k_z$ we may consider the second denominator as
unity. It is interesting that in this model the spectral density $S_u(\omega)$
does not depend on $l$.

This spectral density may be recalculated to the spectral density of equivalent
fluctuations of surface displacement to compare it with other known sources 
of noise:
\begin{eqnarray}
S_{x,\ \beta}^{TD}(\omega)&=&
	\frac {\sqrt{2}\,\beta_{eff}^2\lambda^2\kappa T^2}
		{\pi r_0^2 \sqrt{\omega\rho C\lambda^*}},
\end{eqnarray}

\begin{figure}
\epsfbox{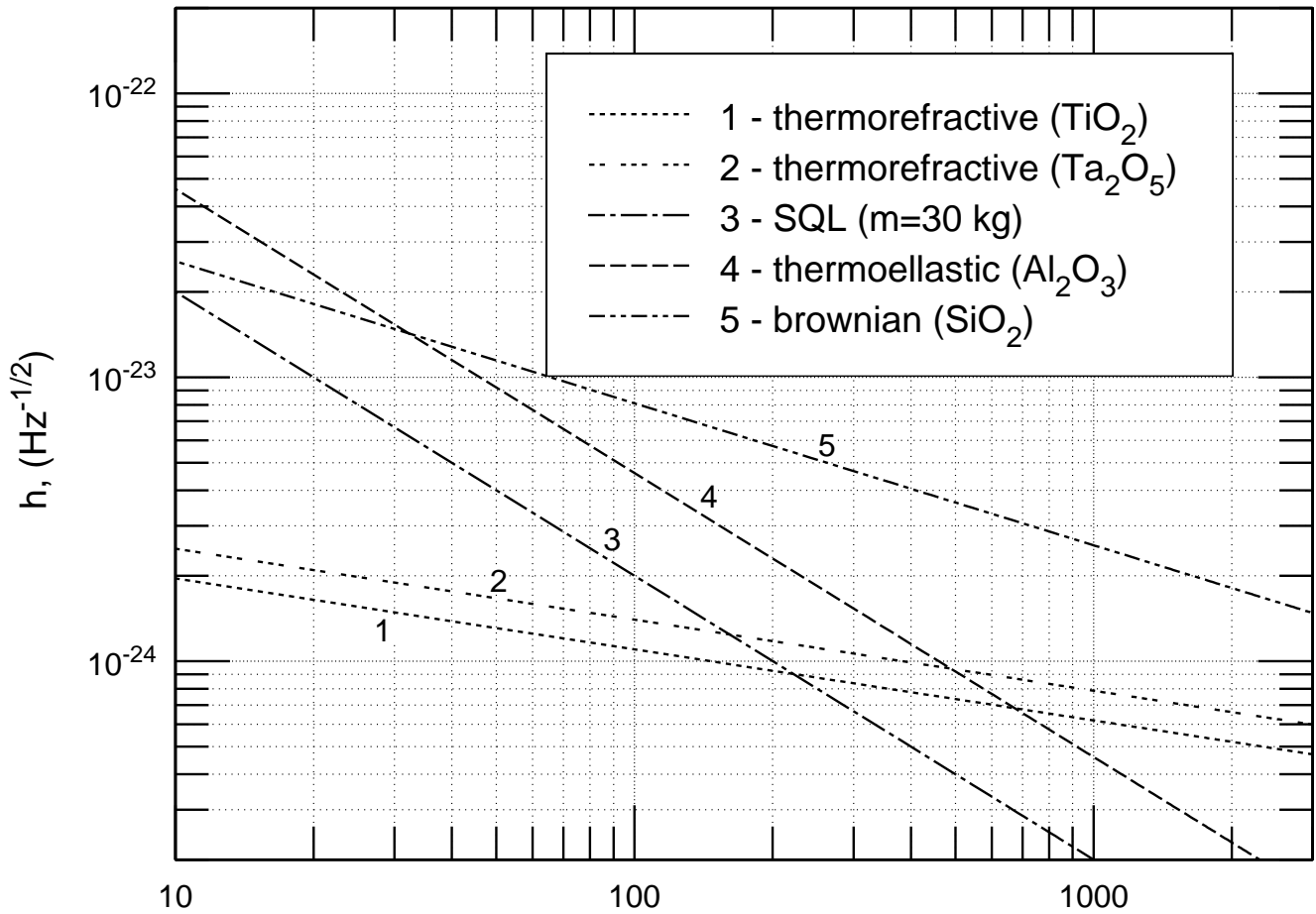}
\caption{Comparison of SQL-limited sensitivity with different sources of noise
in gravitational wave antennae: thermo-refractive, Brownian
(dominating in fused silica mirrors) and thermo-elastic (dominating
in sapphire mirrors).}
\label{fig1}
\end{figure}

\section{Numerical estimates}

For the numerical estimates we assumed that the multilayer coating consists
from alternating pairs of layers: \marginpar{*} 
$TiO_2$ ($n_1=2.2$) and $SiO_2$ ($n_2=1.45$), or 
$Ta_2O_5$ ($n_1=2.2$) and $SiO_2$ ($n_2=1.45$). The values of
$\beta$ for $TiO_2$ and for $Ta_2O_5$ were found in \cite{flory}. 

We want  now to compare the thermo-refractive fluctuations with
thermoelastic noise (\ref{tespot}) and noise associated with the mirrors' 
material losses described in the model of structural damping\cite{saulson} (we
denote it as Brownian motion of the surface). In this model the angle of
losses $\phi$ does not depend on frequency and for its spectral density 
the following formula is valid for infinite test mass \cite{french,bgv,kip}:
\begin{eqnarray}
\label{SDSD}
S^{\rm B}_{x}(\omega)&\simeq&\frac {4\kappa T}{\omega}\,
	\frac{(1-\sigma^2)}{\sqrt{2\pi}  E r_o}\,\phi,
\end{eqnarray}
where $E$ is Young modulus, and $\sigma$ is Poison ratio.

The spectrsal sensitivity of gravitational wave antenna to the perturbation of metric $h(\omega)$
may be recalculated from noise spectral density of displacement $x$
using the following formula:
\begin{equation}
h(\omega)=\frac{
	\sqrt{2(S_{x,\, r_{01}}(\omega)+ S_{x,\, r_{02}}(\omega))}
	}{L},
\end{equation}
where we used the fact that antenna has two arms (with length $L$) with 
two mirrors the fluctuations on which are averaged over the radii $r_{01}$
and $r_{02}$.

The LIGO-II antenna will approach the level of SQL, so we also
compare the noise limited sensitivity to this limit in spectral form 
\cite{thorne}:
\begin{equation}
h_{\rm SQL}(\omega)=\sqrt{\frac{8\hbar}{m\omega^2L^2}}.
\end{equation}

For the calculations we used the set of parameters given 
in Appendix \ref{B} (the same as in \cite{bgv}) plus \cite{flory}
\marginpar{*}
\begin{eqnarray*}
   r_{01}&=&3.6/\sqrt{2}\  \mbox{cm},\quad r_{02}=4.6/\sqrt{2}\  \mbox{cm}, \\
n_1&=&2.2,\qquad \beta_2=4\cdot 10^{-5}\ \mbox{K}^{-1} \quad (TiO_2),\\
n_1&=&2.2,\qquad \beta_2=6\cdot 10^{-5}\ \mbox{K}^{-1} \quad (Ta_2O_5),\\
  n_2&=&1.45,\qquad \beta_1=1.5\cdot 10^{-5}\ \mbox{K}^{-1} \quad (Si O_2),\\
\end{eqnarray*}
We used figures from \cite{flory} for ion plating method only, for other methods of
deposition the value of $\beta$ may be two times larger.
In figure 1 we plot all previously known noises \cite{bgv} together with the
new one.\marginpar{*} 
We see that thermorefractive noise limitation is close to SQL for the
frequences near $200$~Hz.

\section*{Conclusion}

Summing up, we may say that thermo-refractive effect is not small and it must be seriously considered 
in interferometric gravitational antennae (projects LIGO-II and 
especially LIGO-III, where overcomming of the SQL is planned).
It is also important that this effect depends slower on the radii of the
beam-spots than thermo-elastic noise and thus may become dominating for
larger $r_0$ planned in LIGO-II and LIGO-III.

\section*{Acknowledgments}
This material is
based upon work supported by the National Science Foundation under Grant
No. PHY9800097, the Russian Foundation for Basic Research Grant 
\#96-15-96780, and the Russian Ministry of Science and Technology.

\appendix

\section{}\label{A}

In this appendix we give the calculation of coefficient of reflection of light
wave from multilayer coatings consisting of infinite sequences of pairs of 
quarter-wavelength dielectrical layers $n_1$ and $n_2$.

Let the refraction index of odd layers fluctuates on $\Delta n_1$ and the
refraction index of even layers on $\Delta n_2$. One may reformulate this
problem into the problem for distributed long line \cite{solimeno}.
The equivalent impedance $Z$ of this system of layers may be deduced
using the following statement: the addition of two layers does not change the
value of $Z$. 

Voltage  $V_2$ and current $I_2$ at the end of second layer 
may be found from input voltage $V_0$ and current $I_0$ using transformation
matrix $M$ (\cite{solimeno}, formula (3.9.27)):
\begin{eqnarray*}
\left(\begin{array}{c}
	V_2\\ I_2	
	\end{array}\right)&=& M\times 
	\left(\begin{array}{c}
	V_0\\ I_0	
	\end{array}\right), \quad 
	M=\left(\begin{array}{cc}
	M_{11}& M_{12}\\ 
	M_{21}& M_{22}	
	\end{array}\right),\\
M_{11}&=& \cos \phi_1 \cos \phi_2 -\frac{n_1}{n_2}\sin \phi_1 \sin \phi_2,\\ 
M_{12}&=&-i\left(\frac{\sin \phi_1 \cos \phi_2}{n_1}+
		\frac{\sin \phi_2 \cos \phi_1}{n_2}\right),\\
M_{21}&=&	-i\left(n_2\sin \phi_2 \cos \phi_1+
		n_1\sin \phi_1 \cos \phi_2\right),\\
M_{22}&=&\cos \phi_1 \cos \phi_2 -\frac{n_2}{n_1}\sin \phi_1 \sin \phi_2,\\	
M&\simeq&\left(\begin{array}{cc}
	-\frac{n_1}{n_2} &
	i\left(\frac{ \varphi_2}{n_1}+ \frac{ \varphi_1}{n_2}\right)\\
	i\left(n_2 \varphi_1+ n_1\varphi_2\right)&
	 -\frac{n_2}{n_1}	
	\end{array}\right)
\end{eqnarray*}
Here we take into account that for quarter-wavelength layers $\phi_1=
\pi/2 +\varphi_1, \quad \phi_2= \pi/2 +\varphi_2$ and therefore one 
may use approximation  
$\sin \phi_1 \simeq 1,\quad \sin \phi_2 \simeq 1,\quad 
\cos \phi_1\simeq -\varphi_1, \quad  \cos \phi_2\simeq -\varphi_2$.

Now we put that $I_0=YV_0$ and $I_2=YV_2$ ($Y=1/Z$ is generalized
conductivity of the sequence of layers) and obtain two equations:
\begin{eqnarray}
V_2&=& V_0\left(-\frac{n_1}{n_2} +
	iY\left(\frac{ \varphi_2}{n_1}+ \frac{ \varphi_1}{n_2}\right) \right),\\
YV_2&=&V_0\left(i\left(n_2 \varphi_1+ n_1\varphi_2\right)
	-Y\,\frac{n_2}{n_1}\right).
\end{eqnarray}
Solving these equations we find conductivity $Y$ and reflection
coefficient $K$:
\begin{eqnarray}
Y&\simeq& -i\, \frac{n_1n_2}{n_1^2-n_2^2}\,
	\left(n_2\varphi_1 +n_1\varphi_2\right),\\
K&=&\frac{Y-1}{Y+1}\simeq -1 -2i\, \frac{n_1n_2}{n_1^2-n_2^2}\,
	\left(n_2\varphi_1 +n_1\varphi_2\right)
\end{eqnarray}
From this point it is easy to obtain (\ref{deltax},\ref{betaeff}), assuming 
$$ \varphi_1=\frac{\pi}{2}\, \frac{\Delta n_1}{n_1},\qquad
	\varphi_2=\frac{\pi}{2}\, \frac{\Delta n_2}{n_2}
$$

\section{Parameters}\label{B}

\begin{eqnarray*}
\omega&=&2\pi\times 100\ \mbox{s}^{-1},\quad
	T=300\ \mbox{K}, \nonumber\\
m&=&=3\times 10^4\ \mbox{ g},\quad \lambda = 1.06 \ \mu ,\quad 
	L=4\times 10^5\ \mbox{cm}; \label{parameter}; \\
&&\mbox{Fused silica:}\\
\quad \alpha&=&5.5\times  10^{-7}\ \mbox{K}^{-1}, \ \
	\lambda^*=1.4\times  10^5\ \frac{\mbox{erg}}{\mbox{cm s K}},
	\label{silica}\\
\rho&=&2.2\ \frac{\mbox{g}}{\mbox{cm}^3}, \ \
	C=6.7\times  10^6\ \frac{\mbox{erg}}{\mbox{g K}},
	\nonumber \\
E&=&7.2\times 10^{11}\frac{\mbox{erg}}{\mbox{cm}^3},  \ \
	\sigma=0.17,\ \
	\phi=5\times10^{-8}; \nonumber \\
&&\mbox{Sapphire:}\\
\quad \alpha&=&5.0\times  10^{-6}\ \mbox{K}^{-1}, \ \
	\lambda^*=4.0\times  10^6\ \frac{\mbox{erg}}{\mbox{cm s K}},
	\label{saphire}\\
\rho&=&4.0\ \frac{\mbox{g}}{\mbox{cm}^3},\ \
	C=7.9\times  10^6\ \frac{\mbox{erg}}{\mbox{g K}},
	\nonumber\\
E&=&4\times 10^{12}\,\frac{\mbox{erg}}{\mbox{cm}^3},\ \
	\sigma=0.29,\ \
	\phi=3\times10^{-9}.
\end{eqnarray*}

\end{document}